\documentclass[12pt]{article}
\usepackage{epsfig,amsmath,amsfonts,amssymb,amstext,afterpage,psfrag,%slashbox
}
\usepackage{slashed}
\usepackage[square,comma,sort&compress,numbers]{natbib}
\allowdisplaybreaks
\newdimen\TW
\usepackage{color}
\definecolor{light-gray}{gray}{0.9}
\definecolor{dark-gray}{gray}{0.7}
\usepackage{tikz}

\long\def\symbolfootnote[#1]#2{\begingroup%
\def\thefootnote{\fnsymbol{footnote}}\footnote[#1]{#2}\endgroup}

\def\l{\langle}
\def\r{\rangle}

\def\spose#1{\hbox to 0pt{#1\hss}}
\def\lsim{\mathrel{\spose{\lower 3pt\hbox{$\mathchar"218$}}
 \raise 2.0pt\hbox{$\mathchar"13C$}}}
\def\gsim{\mathrel{\spose{\lower 3pt\hbox{$\mathchar"218$}}
 \raise 2.0pt\hbox{$\mathchar"13E$}}}

\catcode`@=11
\def\@citex[#1]#2{%
  \if@filesw\immediate\write\@auxout{\string\citation{#2}}\fi
  \def\@citea{}\@cite{\@for\@citeb:=#2\do
    {\@citea\def\@citea{,\penalty\@m}\@ifundefined
      {b@\@citeb}{{\bf ?}\@warning
{Citation `\@citeb' on page \thepage \space undefined}}%
      \hbox{\csname b@\@citeb\endcsname}}}{#1}}
\def\citer{\@ifnextchar [{\@tempswatrue\@citexr}{\@tempswafalse\@citexr[]}}
  \def\@citexr[#1]#2{%
    \if@filesw\immediate\write\@auxout{\string\citation{#2}}\fi
    \def\@citea{}\@cite{\@for\@citeb:=#2\do
      {\@citea\def\@citea{--\penalty\@m}\@ifundefined
{b@\@citeb}{{\bf ?}\@warning
{Citation `\@citeb' on page \thepage \space undefined}}%
\hbox{\csname b@\@citeb\endcsname}}}{#1}}

\newcommand{\ch}{|h|}

\begin{document}

\begin{titlepage}

\begin{flushright}
{\small
LMU-ASC~49/14\\ 
FLAVOUR(267104)-ERC-77\\
December 2014\\
%Draft \today
%hep-ph/yymmnnn
}
\end{flushright}

\vspace{0.5cm}
\begin{center}
{\Large\bf \boldmath                                               
A Systematic Approach to the SILH Lagrangian
%\vspace*{0.3cm}                                                            
\unboldmath}
\end{center}

\vspace{0.5cm}
\begin{center}
{\sc Gerhard Buchalla$^1$, Oscar Cat\`a$^{1,2,3}$ and Claudius Krause$^1$} 
\end{center}

\vspace*{0.4cm}

\begin{center}
$^1$Ludwig-Maximilians-Universit\"at M\"unchen, Fakult\"at f\"ur Physik,\\
Arnold Sommerfeld Center for Theoretical Physics, 
D--80333 M\"unchen, Germany\\
\vspace*{0.2cm}
$^2$TUM-IAS, Lichtenbergstr. 2a, D--85748 Garching, Germany\\
\vspace*{0.2cm}
$^3$Physik Department, TUM, D--85748 Garching, Germany
\end{center}

\vspace{1.5cm}
\begin{abstract}
\vspace{0.2cm}\noindent
We consider the electroweak chiral Lagrangian, including a light
scalar boson, in the limit of small $\xi=v^2/f^2$.
Here $v$ is the electroweak scale and $f$ is the corresponding scale of
the new strong dynamics. We show how the conventional SILH Lagrangian, 
defined as the effective theory of a strongly-interacting light Higgs 
(SILH) to first order in $\xi$, can be obtained as a limiting case 
of the complete electroweak chiral Lagrangian.
The approach presented here ensures the completeness of the operator basis 
at the considered order, it clarifies the systematics of the effective 
Lagrangian, guarantees a consistent and unambiguous power counting, and
it shows how the generalization of the effective field theory 
to higher orders in $\xi$ has to be performed.
We point out that terms of order $\xi^2$, which are usually not included
in the SILH Lagrangian, are parametrically larger than terms of order
$\xi/16\pi^2$ that are retained, as long as $\xi\gsim 1/16\pi^2$.
Conceptual issues such as custodial symmetry and its breaking
are also discussed. For illustration, the minimal composite 
Higgs model based on the coset $SO(5)/SO(4)$ is considered at 
next-to-leading order in the chiral expansion. It is shown how the
effective Lagrangian for this model is contained as a special case in
the electroweak chiral Lagrangian based on $SU(2)_L\otimes SU(2)_R/SU(2)_V$.

\end{abstract}

\vfill

\end{titlepage}

%%%%%%%%%%%%%%%%%%%%%%%%%%%%%%%%%%%%%%%%%%%%%%%%%%%%%%%%%%%%%%%%%
%   Introduction
%%%%%%%%%%%%%%%%%%%%%%%%%%%%%%%%%%%%%%%%%%%%%%%%%%%%%%%%%%%%%%%%%
\section{Introduction}
\label{sec:intro}

After the discovery of the Higgs-like boson at the LHC
\cite{Aad:2012tfa,Chatrchyan:2012ufa,ATLAS:2013sla,CMS:yva,Aad:2013xqa}
the understanding of its 
precise role in electroweak symmetry breaking has become the prime topic
in particle physics. A general approach, independent of any specific 
high-energy model, is provided by the effective field-theory (EFT) method.
The motivation for using EFT is reinforced by the absence (so far) of 
evidence for new particles below the TeV energy scale.
To be fully general and to account for the possibility of Higgs compositeness,
the electroweak chiral Lagrangian~\cite{Appelquist:1980vg} including a light 
Higgs~\cite{Feruglio:1992wf,Buchalla:2013eza,Buchalla:2012qq,Buchalla:2013rka}
should be employed.

A widely used effective description of a light (pseudo-Goldstone) composite
Higgs particle is the Lagrangian of the strongly-interacting light Higgs
(SILH) \cite{Giudice:2007fh,Contino:2013kra}.
By definition, this low-energy Lagrangian is constructed under the 
additional assumption that the electroweak scale $v$ is parametrically 
smaller than the corresponding scale of the strong dynamics $f$, and the
Lagrangian is restricted to terms of at most first order in $\xi=v^2/f^2$.
For clarity we will therefore define the term {\it SILH Lagrangian} as the
effective theory of a light composite Higgs particle through linear
order in $\xi$. Without this restriction, we will use the
expression {\it electroweak chiral Lagrangian} instead.

As will be explained in detail below, the usual derivation of the SILH
Lagrangian appears unsatisfactory. An important point is that the 
power-counting rules postulated in \cite{Giudice:2007fh} are based on naive 
dimensional analysis (NDA) \cite{Manohar:1983md,Georgi:1992dw}, which are 
not valid in general \cite{Buchalla:2013eza} in their usual formulation.
As a consequence, the counting rules of \cite{Giudice:2007fh} are not fully
consistent and lead to ambiguities in the estimate of the Lagrangian
coefficients.
A systematic derivation of the SILH Lagrangian can be given starting from
the electroweak chiral Lagrangian with a light Higgs.
This derivation is the main subject of this paper. In addition to providing
a complete SILH-type Lagrangian, we elaborate on further conceptual issues
of relevance to phenomenology. Although we agree with many results
incorporated in the traditional SILH Lagrangian, we find some notable
differences, which we discuss. 

Some of the differences between the traditional formulation and the
present approach may be seen as reflecting two different perspectives 
on effective field theory, which we might refer to as the `top-down' and the
`bottom-up' point of view \cite{Georgi:1994qn}. On the one hand, 
in the top-down applications, effective field theory can be used
to construct a low-energy approximation of a given theory, or a certain
class of theories, at high energies. A typical example is the
derivation of low-energy effective Lagrangians for the weak interactions
of the Standard Model (SM), or one of its extensions. 
In this case the high-energy 
theory is known and the EFT is used as a systematic tool to simplify
the theory in the energy regime of interest. On the other hand, following 
the bottom-up approach, a low-energy EFT can be constructed from the relevant
light degrees of freedom, based on the appropriate symmetries and a consistent
power counting, without specifying any details of the high-energy completion.
An example is the chiral perturbation theory of pions, where the
theory at high energies, QCD, is known in principle, but the nonperturbative
hadronic dynamics makes it intractable in practice.
Another example is the renormalizable Standard Model itself, which
can be extended by operators of higher dimension to account in full generality 
for the unknown physics in the UV.
Even though a top-down construction may capture most of the essential
features in the low-energy Lagrangian, it is clear that
only the bottom-up framework will guarantee a fully general EFT.

We emphasize that our starting point, the electroweak chiral Lagrangian,
is formulated as a bottom-up EFT in this sense, whereas the 
traditional SILH Lagrangian might be rather considered as following
a top-down approach with a class of composite Higgs models in mind. In
this work we will show how the SILH Lagrangian can be consistently derived
from a model-independent bottom-up perspective.

The remainder of this paper is organized as follows. In Section~\ref{sec:csilh} 
we revisit the original SILH Lagrangian as given 
in~\cite{Giudice:2007fh} and discuss a number of issues related to its power 
counting. 
Section~\ref{sec:lchxi} outlines the systematics of the electroweak chiral
Lagrangian in the limit of small $\xi$ and clarifies the connection
between a dimensional expansion (in powers of $\xi$) and the chiral
expansion (in the number of loops).
In Section~\ref{sec:silh} we derive the SILH Lagrangian, identified 
as the ${\cal{O}}(\xi)$ expansion of the electroweak chiral 
Lagrangian. Comments on custodial symmetry and its 
breaking through spurions are given in Section~\ref{sec:custodial}. As
a concrete illustration, in Section~\ref{sec:so5nlo} we discuss the basis 
of bosonic NLO operators for the $SO(5)/SO(4)$ model. 
Conclusions are given in Section~\ref{sec:concl} while 
technical details are collected in an Appendix.      

%%%%%%%%%%%%%%%%%%%%%%%%%%%%%%%%%%%%%%%%%%%%%%%%%%%%%%%%%%%%%%%%%
%   SILH comments
%%%%%%%%%%%%%%%%%%%%%%%%%%%%%%%%%%%%%%%%%%%%%%%%%%%%%%%%%%%%%%%%%
\section{Comments on the SILH Lagrangian}
\label{sec:csilh}

The construction of the electroweak chiral Lagrangian as an EFT 
requires a power-counting prescription in order to be well defined. 
As discussed in~\cite{Buchalla:2013eza}, the electroweak chiral 
Lagrangian mixes weakly-coupled and strongly-coupled interactions, which in 
isolation have a very different power counting. The strategy followed 
in~\cite{Buchalla:2013eza,Buchalla:2012qq,Buchalla:2013rka}
was to define a power counting such that NLO counterterms account for 
all the (superficial) divergences coming from the one-loop diagrams built from 
the leading-order Lagrangian.\footnote{Contrary to the chiral Lagrangian in 
the strong sector, the loop expansion for the electroweak chiral Lagrangian 
cannot be cast as a derivative expansion. Quite generally, a derivative 
expansion is valid only when the field content of the theory is restricted 
to Goldstone fields, but fails when other fields are present. The Yukawa 
interactions, for instance, clearly cannot be accounted for in terms of a 
pure derivative expansion.} 
Such a power counting is thus based on the infrared structure of the theory and 
in this sense it is the most general one. In particular, it allows us to 
identify the natural size of the coefficients associated with each order in 
the EFT expansion. 

Let us briefly summarize the basic assumptions and properties of this
framework.
\begin{itemize}
\item
The Goldstone bosons of electroweak symmetry breaking and the light Higgs 
are treated, in general, as part of a new strong
dynamics, to which they are coupled with a strength of ${\cal O}(4\pi)$. The 
scale of the new dynamics is given by the Goldstone-boson decay constant $f$.
\item
The transverse gauge bosons and the fermions of the Standard Model are
weakly coupled among themselves and to the strong sector, that is with
couplings of ${\cal O}(1)$.
\item
The general effective theory for the light fields mentioned above
(the fields of the SM) is an electroweak chiral Lagrangian. This theory is
nonrenormalizable and is valid below a cut-off\footnote{If new states with mass
of order $f\gg v$ are present, the chiral Lagrangian with SM fields will
only be valid up to this scale $f$.}  
$\Lambda=4\pi f$. The terms in the Lagrangian are organized as a loop 
expansion, which is equivalent to a counting of terms according to their 
chiral dimension \cite{Buchalla:2013eza,Urech:1994hd,Nyffeler:1999ap}.
The assignment of chiral dimensions to fields and couplings is
$0$ for Higgs, Goldstone and gauge fields, $1/2$ for
fermions, and $1$ for derivatives and weak couplings (gauge or Yukawa).
\item
In full generality, the electroweak scale $v$ and the scale $f$ can be taken
to be of the same order, $\xi\equiv v^2/f^2={\cal O}(1)$.
An expansion in $\xi$ can be performed for $\xi\ll 1$.
In this case a counting by canonical dimension is recovered.
The new dynamics decouples in the limit $\xi\to 0$.
\end{itemize}

Knowledge or partial knowledge of non-infrared physics, {\it{i.e.}}, extra 
symmetries or additional particle content around or beyond the cutoff 
$\Lambda=4\pi f$, refine the power-counting estimate and allow for additional 
information on the size of the operator coefficients. However, one should 
keep in mind that incorporating UV information goes beyond the 
EFT power counting and introduces some degree of model dependence.

The traditional SILH Lagrangian \cite{Giudice:2007fh} is constructed by 
assuming a set of (infrared) power-counting rules based on 
NDA~\cite{Manohar:1983md,Georgi:1992dw} supplemented with information on 
the UV completion. Specifically, the UV completion is assumed to contain a 
heavy vector with $m_V\lesssim 4\pi f=\Lambda$, implemented as a gauge field 
of some hidden local symmetry (HLS)~\cite{Bando:1984ej,Bando:1987br}. 
The NDA rules for operator building in \cite{Giudice:2007fh} are only defined
relative to the leading-order (SM) Lagrangian, rather than in an absolute
sense, and read: 
(i) extra powers of the Higgs doublet $H$ receive a suppression by $1/f$; 
(ii) SM gauge fields and derivatives receive a $1/m_V$ suppression.

With these assumptions, the NLO Lagrangian is written as~\cite{Giudice:2007fh}:
\begin{eqnarray}\label{SILH}
{\cal{L}}_{SILH} &=&
\frac{c_H}{2f^2}\partial^{\mu}(H^{\dagger}H)\partial_{\mu}(H^{\dagger}H)
-\frac{c_6\lambda}{f^2}(H^{\dagger}H)^3+
\left(\frac{c_y y_f}{f^2}H^{\dagger}H{\bar{f}}_LHf_R+{\mathrm{h.c.}}\right)
\nonumber\\
&+&\frac{c_T}{2f^2}(H^{\dagger}\!\!\stackrel{\longleftrightarrow}{D^{\mu}}\!\!H)
(H^{\dagger}\!\!\stackrel{\longleftrightarrow}{D_{\mu}}\!\!H)\nonumber\\
&+& ig\frac{c_W}{2m_V^2}
(H^{\dagger}\!\!\stackrel{\longleftrightarrow}{D_{\mu}^a}\!\!H)(D^{\nu}W_{\mu\nu})^a
+ig^{\prime}\frac{c_B}{2m_V^2}
(H^{\dagger}\!\!\stackrel{\longleftrightarrow}{D_{\mu}}\!\!H)
(\partial^{\nu}B_{\mu\nu})\nonumber\\
&+&ig\frac{c_{HW}}{(4\pi f)^2}D^{\mu}H^{\dagger}W_{\mu\nu}{D^{\nu}}H
+ig^{\prime}\frac{c_{HB}}{(4\pi f)^2}D^{\mu}H^{\dagger}{D^{\nu}}HB_{\mu\nu}\nonumber\\
&+&g^{\prime 2}\frac{c_{\gamma}}{(4\pi f)^2}\frac{g^2}{g_V^2}
H^{\dagger}HB_{\mu\nu}B^{\mu\nu}
+g_s^2\frac{c_g}{(4\pi f)^2}\frac{y_t^2}{g_V^2}H^{\dagger}HG_{\mu\nu}^aG^{\mu\nu a}
\nonumber\\
&-&g^2\frac{c_{2W}}{2(g_V m_V)^2}D_{\mu}W^{\mu\nu a}D^{\rho}W_{\rho\nu}^a
-g^{\prime 2}\frac{c_{2B}}{2(g_V m_V)^2}\partial_{\mu}B^{\mu\nu}\partial^{\rho}B_{\rho\nu}
\nonumber\\
&-&g_s^2\frac{c_{2G}}{2(g_V m_V)^2}D_{\mu}G^{\mu\nu A}D^{\rho}G_{\rho\nu}^A\nonumber\\
&+&g^3\frac{c_{3W}}{(4\pi m_V)^2}\langle W^{\mu\nu}W_{\nu\rho}W_{\mu}^{\rho}\rangle
+g_s^3\frac{c_{3G}}{(4\pi m_V)^2} \langle G^{\mu\nu}G_{\nu\rho}G_{\mu}^{\rho}\rangle
\end{eqnarray}
where $\langle\ldots\rangle$ denotes the trace.

The first two lines collect the operators that are sensitive to the 
breaking scale $f$, whereas the remaining lines gather operators  
generated either by tree-level resonance exchange or at one loop.

If one is aiming at a general description of strongly-coupled EWSB scenarios, 
the previous setting is unsatisfactory for a number of reasons. The first one 
is the absence of fermionic operators, which were recently included 
in~\cite{Contino:2013kra}. In this paper we will focus instead on issues 
that mostly affect the foundations and systematics of the SILH construction.

The need for a more systematic power counting can be seen from the fact that 
the NDA rules given in~\cite{Giudice:2007fh} lead to ambiguities. As a simple 
example, consider the NLO operator $H^{\dagger}HB_{\mu\nu}B^{\mu\nu}$. This 
operator could be built by either (i) applying the first rule to the gauge 
kinetic term; (ii) applying the second rule to the Higgs potential; or (iii) 
applying the second rule to the Higgs kinetic term. 
The size of the corresponding coefficient would be, respectively, 
${\cal{O}}(1/f^2)$, ${\cal{O}}(f^2/m_V^4)$ and ${\cal{O}}(1/m^2_V)$. 
The right counting is the latter, which follows without ambiguities from the 
rules given in~\cite{Buchalla:2013eza,Buchalla:2013rka}.
Another example is given by the operator 
$(H^{\dagger}\!\!\stackrel{\longleftrightarrow}{D^{\mu}}\!\!H)^{2}$, which 
corresponds to the $T$ parameter. It can be built with rule (i) applied to 
the Higgs kinetic term or with rule (ii) applied to the Higgs quartic 
interaction. The size of the coefficients would then be of ${\cal{O}}(1/f^2)$ 
or ${\cal{O}}(1/m^2_V)$, respectively. In this case, additional dynamical 
assumptions are needed to decide between the different possibilities. A more 
detailed discussion of this operator is given at the end of 
Section \ref{sec:silh}. 

If Eq.~(\ref{SILH}) is to describe a consistent EFT, then the scales 
$\Lambda = 4\pi f\approx m_V\approx g_V f$ may all be identified for 
the purpose of power counting: numerical differences in the size of the 
coefficients are expressed 
in any case through differences in the ${\cal O}(1)$ parameters $c_i$. 
This follows from rather general principles: unless 
$m_V \sim {\cal{O}}(\Lambda)$, the naturalness of the EFT would be upset. 
This is actually one of the conditions to have a natural and predictive 
strongly-coupled EFT (like chiral perturbation theory). 

Based on the previous point, it is apparent that the distinction between 
tree-level ($1/m^2_V$) {\it{vs.}} loop-suppressed operators ($1/16\pi^2 f^2$) 
turns out to be of little 
numerical relevance. However, such a classification is also parametrically 
misleading: which operators can be generated at tree level and which at one 
loop mostly depends on the UV completion one adopts. This point 
has already been discussed in detail in~\cite{Jenkins:2013fya}.

The pattern displayed in (\ref{SILH}) is specific for a UV scenario with 
vector mesons implemented {\it{\`a la}} HLS~\cite{Bando:1984ej,Bando:1987br}. 
However, there is no compelling reason why such a pattern should be expected. 
The best counterexample is provided by QCD itself at low energies 
(see~\cite{Jenkins:2013fya} for similar considerations). 
Within the chiral expansion, the NLO operators that involve gauge fields can 
be written as \cite{Gasser:1983yg}
\begin{align}
{\cal{L}}_{\chi}^{(4)}&=L_9^L\langle F^{\mu\nu}_LD_{\mu}UD_{\nu}U^{\dagger}\rangle+
L_9^R\langle F^{\mu\nu}_RD_{\mu}U^{\dagger}D_{\nu}U\rangle\nonumber\\
&+L_{10}\langle F^{\mu\nu}_LUF_{R\mu\nu}U^{\dagger}\rangle+
H_1^L\langle F^{\mu\nu}_LF_{L\mu\nu}\rangle+H_1^R\langle F^{\mu\nu}_RF_{R\mu\nu}\rangle
\end{align}  
where $F_{\mu\nu}^{L,R}$ are generic (non-Abelian) external sources. Since QCD 
has a global $SU(2)_L\otimes SU(2)_R$ symmetry broken down to $SU(2)_V$, 
$L_9^L=L_9^R\equiv L_9$ and $H_1^L=H_1^R\equiv H_1$. However, for comparison 
purposes, it is useful to formally distinguish the left- and right-handed 
parts. By inspection one can then see that ${\cal{O}}_{HW}$, 
${\cal{O}}_{HB}$ and ${\cal{O}}_{\gamma}$ are in correspondence with 
${\cal{O}}_{9}^L$, ${\cal{O}}_{9}^R$ and ${\cal{O}}_1^R$, respectively. 
${\cal{O}}_{W}$ and ${\cal{O}}_B$, in turn, can be rewritten as linear 
combinations of the previous operators and, additionally, 
$H^{\dagger}HW_{\mu\nu}^aW^{\mu\nu a}$ and $H^{\dagger}W_{\mu\nu}HB^{\mu\nu}$, which 
correspond to ${\cal{O}}_1^L$ and ${\cal{O}}_{10}$. Schematically, 
\begin{align}
{\cal{O}}_{HW,HB}\sim {\cal{O}}_9^{L,R};\qquad\qquad {\cal{O}}_{\gamma}\sim 
{\cal{O}}_1^R;\qquad\qquad {\cal{O}}_{W,B}\to {\cal{O}}_{10},{\cal{O}}_1^L
\end{align}
In QCD, all the previous operators are experimentally of the same order 
$\sim f_{\pi}^2/\Lambda^2$, they all can be generated by tree-level resonance 
exchange, and they all are ${\cal{O}}(N_c)$, with no combinations of them 
being suppressed, {\it{i.e.}}, ${\cal{O}}(1)$ \cite{Ecker:1988te}. 
Therefore, QCD does not follow the UV pattern assumed 
in~\cite{Giudice:2007fh}. 

A point of phenomenological relevance is the spurion suppression associated 
with $H^{\dagger}HB_{\mu\nu}B^{\mu\nu}$ and $H^{\dagger}HG_{\mu\nu}^AG^{\mu\nu A}$, 
$(g/g_V)^2$ and $(y_t/g_V)^2$, respectively. This issue is closely related to 
shift symmetry and its breaking, and will be discussed in more detail in 
Section~\ref{sec:so5nlo}. The main conclusion is that such a suppression
is not present in general. 

To summarize, the operators collected in~(\ref{SILH}) have a simple counting 
in terms of the breaking scale $f$ and the cutoff scale $4\pi f$ and fall 
into four main classes: 
\begin{itemize}
\item The first line is suppressed only by $1/f^2$, which in the 
electroweak chiral Lagrangian corresponds to LO operators.
\item The second line ($T$-parameter) is superficially of order
$1/f^2$. If custodial symmetry is weakly broken, as it is usually assumed,
the actual coefficient comes with an extra suppression by $1/16\pi^2$.
\item The last line, as it stands,
would correspond to NNLO operators, since effectively they 
are two-loop suppressed, ${\cal{O}}(1/(16\pi^2 f)^2)$. 
\item The remaining operators carry a $1/(4\pi f)^2$ suppression. In the 
electroweak chiral Lagrangian they appear as NLO operators, and as such they 
are generated by LO loop diagrams as well as tree-level resonance exchange, 
in analogy with what happens in QCD. The additional $(g/g_V)^2$ and 
$(y_t/g_V)^2$ factor suppression 
in $H^{\dagger}H B_{\mu\nu}B^{\mu\nu}$ and $H^{\dagger}H G_{\mu\nu}G^{\mu\nu}$ is not
present in a model-independent way.  
\end{itemize}

In the following sections we will substantiate these statements by deriving 
the SILH Lagrangian as a limiting case of the more general electroweak 
chiral Lagrangian.

%%%%%%%%%%%%%%%%%%%%%%%%%%%%%%%%%%%%%%%%%%%%%%%%%%%%%%%%%%%%%%%%%
%   SILH derivation
%%%%%%%%%%%%%%%%%%%%%%%%%%%%%%%%%%%%%%%%%%%%%%%%%%%%%%%%%%%%%%%%%
\section{The electroweak chiral Lagrangian at small $\xi$}
\label{sec:lchxi}

We will next outline the systematics of the effective theory
for standard-model particles with strong dynamics in the Higgs sector.
The basic assumptions for the fields and their couplings have been
summarized at the beginning of Section~\ref{sec:csilh}.

The framework is very general and can be applied to different scenarios.
When the appropriate limits are taken, it covers technicolor-like theories, 
composite-Higgs models, or models with weakly-coupled UV completions. 
To be specific, we will focus on theories with a pseudo-Goldstone Higgs. 
In this case we can typically distinguish
three relevant energy scales: The electroweak scale $v$, the scale $f$
of the symmetry breaking that leads to the Goldstone bosons, and the scale
$\Lambda=4\pi f$, where the low-energy description of this dynamics
is cut off. The three scales imply two possible expansion parameters,
$\xi = v^2/f^2$ and the loop factor $1/(16\pi^2)= f^2/\Lambda^2$. 
 
\begin{figure*}[t]
\begin{center}
\includegraphics[width=8cm]{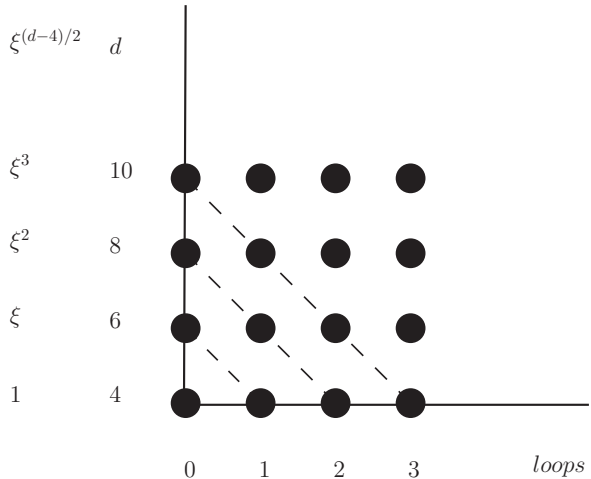}
\end{center}
\caption{Systematics of the effective theory with strong
dynamics in the Higgs sector. The dots indicate operators in
the effective Lagrangian (or terms in a physical amplitude). In general, 
they may be organized both in powers of $\xi=v^2/f^2$ (vertical axis) 
and according to their order $L$ in the loop expansion (horizontal axis).
The latter is equivalent to the chiral dimension $2L+2$.}
\label{fig:loopdim}
\end{figure*}

The resulting picture is sketched in Fig. \ref{fig:loopdim}, where we
plot the powers of $\xi$ on the vertical axis and the loop order on the
horizontal. 
The dots indicate, schematically, (classes of) operators in
the effective Lagrangian or, alternatively, terms in a physical
amplitude.

Without expanding in $\xi$, the effective theory takes the form of
a loop expansion as in the usual chiral Lagrangians \cite{Weinberg:1978kz}. 
This amounts to proceeding from left to right in Fig. \ref{fig:loopdim}, 
order by order in the loop expansion, resumming at each order all terms 
along the vertical axis. 

Alternatively, the expansion may be organized in powers of $\xi$,
proceeding from bottom to top of Fig. \ref{fig:loopdim} and including,
in principle, at each power of $\xi$ terms of arbitrary order in the 
loop expansion. This scheme corresponds to the conventional expansion
of the effective theory in terms of the canonical dimension $d$ of
operators, where the power of $\xi$ is given by $(d-4)/2$.
Since the dimensional expansion requires only a hierarchy between $v$ and
the new-physics scale $f$, $\xi\ll 1$, it is not restricted to
the pseudo-Goldstone Higgs scenarios we are focussing on here.

We emphasize that these observations clarify the relation between
an effective theory organized by canonical dimension and
the electroweak chiral Lagrangian organized as a loop expansion:
The former is constructed row by row, the latter column by column
from the terms in Fig. \ref{fig:loopdim}.

We now return specifically to the pseudo-Goldstone Higgs scenario with
a hierarchy between $v$ and $f$. The Higgs sector is assumed to be governed
by strong dynamics. Its effective description at scale $f$ is then
organized in terms of a loop expansion. The electroweak effective Lagrangian
at scale $v$ is further obtained by integrating out the physics at $f$,
which amounts to a dimensional expansion in powers of $\xi$. Therefore,
if $\xi$ is small enough for this expansion to be 
meaningful\footnote{If $v$ is not much smaller than $f$, the $\xi$-expansion
cannot be performed. The resulting EFT is a chiral Lagrangian at scale $f$.
In this case, if new particles with mass of order $f$ should exist, they
would have to be included as additional fields in the EFT. This is of course
possible, but it would go beyond our initial assumption of a standard-model 
particle content.}, the effective theory at $v$ can be considered
as a {\it double expansion} in the number of loops and in powers of $\xi$.
Put differently, the expansion is governed simultaneously by chiral and
canonical dimensions. Nominally taking $\xi$ and $f^2/\Lambda^2$ to be 
of the same order, the effective theory for a pseudo-Goldstone Higgs sector 
then becomes an expansion organized as indicated by the dashed lines 
in Fig. \ref{fig:loopdim}.

We note that the conventional SILH Lagrangian 
\cite{Giudice:2007fh,Contino:2013kra} has been defined as a dimensional 
expansion up to first order in $\xi$, with a further scaling
of the coefficients of dimension-6 operators either as $1/f^2$ or
$1/\Lambda^2$. As discussed in Section~\ref{sec:csilh}, the latter scaling 
essentially reproduces the weighting implied by the loop expansion.
However, at second order in the double expansion only terms of
order $\xi/16\pi^2$ are retained. The terms of order $\xi^2$ are not
included in SILH, which is not justified as long as $\xi$ is at least
of order $1/16\pi^2$. In fact, one typically has $\xi\gg 1/16\pi^2$,
in which case $\xi^2$ is actually more important than $\xi/16\pi^2$.
This holds in spite of the fact that $\xi^2$ terms correspond to
operators of canonical dimension 8.
Such a behaviour may seem unexpected from the point of view of an
effective theory organized primarily in terms of canonical dimensions.
It can be understood, however, as a natural consequence of the power
counting based on chiral dimensions underlying the EFT of a 
strongly-coupled sector.

We remark that the terms of order $\xi^2$ might be included in the
conventional SILH Lagrangian by adding the appropriate operators of
dimension 8 (and leading chiral dimension). 
Alternatively, we may simply choose to work throughout
with the electroweak chiral Lagrangian, which automatically includes
a resummation to all orders in $\xi$.

Finally, we emphasize again the very general nature of the complete
electroweak chiral Lagrangian with a light Higgs as presented
in~\cite{Buchalla:2013rka}. There the Higgs boson is simply described
as an electroweak singlet, coupled to the nonlinearly realized electroweak
Goldstone bosons and the remaining SM fields. While this covers scenarios 
where the Higgs is a pseudo-Goldstone boson from an extended symmetry,
it is not restricted to them. The Higgs particle might e.g. be a dilaton,
or just an ad-hoc singlet, even though this appears unattractive
theoretically. In any case, the chiral Lagrangian framework will allow for
experimental tests with the minimum amount of theoretical bias.

%%%%%%%%%%%%%%%%%%%%%%%%%%%%%%%%%%%%%%%%%%%%%%%%%%%%%%%%%%%%%%%%%
%   SILH derivation
%%%%%%%%%%%%%%%%%%%%%%%%%%%%%%%%%%%%%%%%%%%%%%%%%%%%%%%%%%%%%%%%%
\section{SILH from the electroweak chiral Lagrangian}
\label{sec:silh}

While EFTs of weakly-coupled dynamics are dimensional expansions in powers 
of $1/\Lambda $, EFTs of strongly-coupled dynamics are intrinsically loop 
expansions. As a result, they are expansions in 
$f^2/\Lambda^2 = 1/(16 \pi^{2} )$, which is a reflection of the 
nondecoupling nature of the interactions. Scenarios that incorporate the 
vacuum misalignment mechanism~\cite{Kaplan:1983fs,Dugan:1984hq} 
allow us to describe the transition from the nondecoupling to the decoupling 
regime through the parameter $\xi=v^2/f^2$, such that 
at small $\xi$ one recovers a linear (dimensional) expansion. This means that 
the electroweak effective Lagrangian, which is generally defined as   
\begin{align}
{\cal{L}}_{\chi EW}&={\cal{L}}_{LO}^{(\xi)}+
{\cal{L}}_{NLO}^{(\xi)}+{\cal{O}}\left(\frac{f^4}{\Lambda^4}\right)
\end{align}
with ${\cal L}_{NLO}^{(\xi)}={\cal O}(f^2/\Lambda^2)$, should satisfy
\begin{align}
\lim_{\xi\to0}{\cal{L}}_{\chi EW}&={\cal{L}}_{(0)}+\xi{\cal{L}}_{(1)}+
{\cal{O}}(\xi^2)\equiv{\cal{L}}_{SM}+\xi {\bar {\cal L}}_{SILH}+{\cal{O}}(\xi^2)
\end{align}
It follows that ${\cal{L}}_{SM}={\cal{L}}_{LO}^{(\xi=0)}$, while 
${\cal L}_{SILH} \equiv \xi {\bar {\cal L}}_{SILH}$, with 
\begin{equation}\label{lbarsilh}
{\bar {\cal L}}_{SILH}=\left[ \frac{d}{d\xi} 
({\cal{L}}_{LO}^{(\xi)}+{\cal{L}}^{(\xi)}_{NLO})\right]_{\xi\to 0}
\end{equation}
This non-trivial overlap between LO and NLO operators of the linear and 
non-linear bases has already been discussed in~\cite{Buchalla:2013rka}.  
Dimension-six operators coming from ${\cal{L}}_{LO}$ are suppressed by $1/f^2$, 
whereas dimension-six operators stemming from ${\cal{L}}_{NLO}$ have a 
$1/\Lambda^2$ suppression. The contribution of 
${\cal{L}}_{LO}$ to every order in the $\xi$-expansion can be easily 
understood by noticing that powers of $H^{\dagger}H= (v+h)^2/2$ 
increase canonical dimensions but leave chiral dimensions unaffected. 

The generic dipole ${\bar{\psi}}\sigma_{\mu\nu}X^{\mu\nu}\psi$ 
and triple-field-strength $X_{\mu\nu}X^{\nu\lambda}X_{\lambda}^{\mu}$ operators 
are not required as counterterms of the chiral Lagrangian at NLO.
Concerning their importance in the EFT we remark that
the counting of chiral dimensions is less straightforward than the counting
of canonical dimensions since the number of weak couplings 
(carrying a chiral dimension of 1) is not always obvious from the field
content of a given operator. 
We will next discuss some consequences of this in more detail, considering 
first the chiral Lagrangian at scale $f$, where the physics that has been
integrated out resides at scale $\Lambda$.

The triple-gauge operators $X^3$ have three
derivatives. The gauge fields are weakly coupled to the heavy sector, which
implies the presence of (at least) three gauge couplings. This is true
irrespective of whether the heavy sector itself is governed by weakly or
strongly coupled dynamics. It follows that $g^3 X^3$ has chiral dimension 6
and therefore enters only at NNLO, that is with a double suppression
$\sim 1/(16\pi^2\Lambda^2)$. We remark that this argument generalizes the
corresponding result of \cite{Arzt:1994gp}, obtained for the case of a 
weakly-coupled UV completion. An explicit example for the 
$1/(16\pi^2\Lambda^2)$ scaling of the coefficient of $X^3$ in the context of
a strongly-coupled heavy sector is given by the model discussed 
in \cite{Manohar:2013rga}.
We emphasize that here the scaling of the coefficient does not automatically
follow from the canonical dimension of the operator, which only implies
a factor of $1/\Lambda^2$. On the other hand, the presence of the
additional loop factor $1/(16\pi^2)$ is consistently accounted for through
the counting of chiral dimensions.
The situation may change when new states at the scale $f\gg v$
are integrated out to yield the EFT at the scale $v$. 
In this case, coefficients of order $\xi/16\pi^2$ could arise
for the $X^3$ operators.
Similar comments apply to the dipole operators 
$m_\psi \bar\psi_L\sigma_{\mu\nu}\psi_R\, gX^{\mu\nu}$.

In addition, some of the four Fermi operators are not needed as one-loop 
coun\-ter\-terms. However, they can be generated via tree-level exchange of a 
heavy resonance and are therefore kept at NLO.

Practically, as explained in \cite{Buchalla:2013rka},
the list of operators up to linear order in $\xi$ is obtained by 
taking the full list of dimension-six 
operators in the linear basis~\cite{Grzadkowski:2010es,Buchmuller:1985jz} 
and performing the polar decomposition of the doublet 
\begin{equation}
  \label{eq:1}
  \phi = \frac{v+h}{\sqrt{2}} U \begin{pmatrix}0\\1 \end{pmatrix}
\end{equation}
where $U=\exp(2i\varphi^a T^a/v)$ denotes the Goldstone-boson matrix.

The resulting operators are matched onto the leading and next-to-leading 
operators of the chiral Lagrangian. An important subtlety is worth mentioning, 
which affects the whole matching procedure. Since the linear basis is normally 
expressed in the unbroken phase, while the chiral Lagrangian is written in the 
broken phase, in the former case there are NLO contributions that renormalize 
the LO parameters. The modified operators can be brought back to their 
canonical form by subsequent redefinitions of the fields and couplings. 
Here we will omit details of such redefinitions and present the final results. 

To leading order in chiral dimensions, and to first order in $\xi$, 
the SM effective Lagrangian can be written in nonlinear notation 
as \cite{Buchalla:2013rka} 
\begin{eqnarray}\label{l2}
{\cal L}_2 &=& -\frac{1}{2} \langle G_{\mu\nu}G^{\mu\nu}\rangle
-\frac{1}{2}\langle W_{\mu\nu}W^{\mu\nu}\rangle 
-\frac{1}{4} B_{\mu\nu}B^{\mu\nu}
+\bar q i\!\not\!\! Dq +\bar l i\!\not\!\! Dl
 +\bar u i\!\not\!\! Du +\bar d i\!\not\!\! Dd +\bar e i\!\not\!\! De 
\nonumber\\
&& +\frac{v^2}{4}\ \l L_\mu L^\mu \r\, \left( 1+F_U(h)\right)
+\frac{1}{2} \partial_\mu h \partial^\mu h - V(h) \nonumber\\
&& - v \left[ \bar q \left( Y_u +
       \sum^3_{n=1} Y^{(n)}_u \left(\frac{h}{v}\right)^n \right) U P_+r 
+ \bar q \left( Y_d + 
     \sum^3_{n=1} Y^{(n)}_d \left(\frac{h}{v}\right)^n \right) U P_-r
  \right. \nonumber\\ 
&& \quad\quad\left. + \bar l \left( Y_e +
   \sum^3_{n=1} Y^{(n)}_e \left(\frac{h}{v}\right)^n \right) U P_-\eta 
+ {\rm h.c.}\right]
\end{eqnarray}
with $L_\mu=i UD_\mu U^\dagger$, $P_\pm = 1/2\pm T_3$, and
\begin{align}
F_U &=(2-a_2)\frac{h}{v} +(1-2a_2)\left(\frac{h}{v}\right)^2 -\frac{4}{3} a_2 
\left(\frac{h}{v}\right)^3-\frac{1}{3} a_2 \left(\frac{h}{v}\right)^4
\end{align}
\begin{align}
V&=\frac{m^2_h}{2} h^2 + \frac{m^2_h v^2}{2}\left[ 
\left( 1+\frac{4}{3}a_1-\frac{3}{2}a_2\right)\left(\frac{h}{v}\right)^3
+\left(\frac{1}{4}+ 2 a_1 - \frac{25}{12}a_2\right)
\left(\frac{h}{v}\right)^4 \right. \nonumber\\ 
&\left. \hspace*{3cm} +(a_1-a_2)\left(\frac{h}{v}\right)^5+
\frac{a_1-a_2}{6}\left(\frac{h}{v}\right)^6\right]
\end{align}
\begin{align}
Y^{(1)}_f&=\left(1-\frac{a_2}{2}\right)Y_f+ 2\bar Y_f,\qquad
Y^{(2)}_f=3 Y^{(3)}_f = -\frac{a_2}{2} Y_f + 3 \bar Y_f,\qquad f=u,d,e
\end{align}
For generality, we have included generic flavor matrices $\bar Y_f$ arising 
at NLO. In scenarios with minimal flavor violation \cite{D'Ambrosio:2002ex}, 
$\bar Y_f \propto Y_f$.

Here $a_1$, $a_2$ and the flavor matrices $\bar Y_d,\ldots$
correspond to the coefficients of the di\-men\-sion-6 operators 
$(\phi^\dagger\phi)^3$, $\partial(\phi^\dagger\phi)\partial(\phi^\dagger\phi)$ 
and $\bar q\phi d \phi^\dagger\phi,\ldots$, respectively. These coefficients
are all of order $\xi$. When they are put to zero, ${\cal L}_2$ reduces to
the renormalizable SM. 

At chiral dimension 4 (NLO) and to order $\xi$ one finds the Lagrangian
\begin{align}\label{eq:7}
\mathcal{L}_4 &= 
-\beta_1 v^2\langle L_{\mu}\tau_L\rangle^2\left(1+\frac{h}{v}\right)^{4} 
-\frac{c_{Xh1}}{4} B_{\mu\nu}B^{\mu\nu}
\left[1-\left(1+\frac{h}{v}\right)^2\right] \nonumber\\
& -\frac{c_{Xh2}}{2}\langle W_{\mu\nu}W^{\mu\nu}\rangle \left[1-
\left(1+\frac{h}{v}\right)^2\right]-
\frac{c_{Xh3}}{2}\langle G_{\mu\nu}G^{\mu\nu}\rangle 
\left[1-\left(1+\frac{h}{v}\right)^2\right]\nonumber\\
&+c_{XU1} gg^{\prime}\langle W_{\mu\nu} \tau_{L}\rangle B^{\mu\nu} 
\left(1+\frac{h}{v}\right)^{2} + 
c_{\psi V7} (\bar{l}\gamma^{\mu}l) 
\langle L_{\mu} \tau_{L}\rangle \left(1+\frac{h}{v}\right)^{2}\nonumber\\
&+ c_{\psi V1} (\bar{q}\gamma^{\mu}q) 
\langle L_{\mu} \tau_{L}\rangle \left(1+\frac{h}{v}\right)^{2}+ 
c_{\psi V10} (\bar{e}\gamma^{\mu}e) 
\langle L_{\mu} \tau_{L}\rangle \left(1+\frac{h}{v}\right)^{2}\nonumber\\
&+ c_{\psi V4} (\bar{u}\gamma^{\mu}u) 
\langle L_{\mu} \tau_{L}\rangle \left(1+\frac{h}{v}\right)^{2}+ 
c_{\psi V5} (\bar{d}\gamma^{\mu}d) 
\langle L_{\mu} \tau_{L}\rangle \left(1+\frac{h}{v}\right)^{2}\nonumber\\
&+c_{\psi V6} (\bar{u}\gamma^{\mu}d)
\langle P_{21}U^{\dagger}L_{\mu}U\rangle \left(1+\frac{h}{v}\right)^{2}+ 
\text{h.c.}\nonumber\\
&+c_{\psi Vq} \mathcal{O}_{q} \left(1+\frac{h}{v}\right)^{2}+
c_{\psi Vl} \mathcal{O}_{l} \left(1+\frac{h}{v}\right)^{2} + 
\mathcal{L}_{\psi^{4}} + \mathcal{L}_{\psi^2 X} + \mathcal{L}_{X^3}
\end{align}
where $\mathcal{L}_{\psi^{4}}$ refers to all baryon-number conserving 
four-fermion operators, $\mathcal{L}_{\psi^2 X}$ to the dipole operators 
${\bar{\psi}}\sigma_{\mu\nu}X^{\mu\nu}\psi$, and $\mathcal{L}_{X^3}$ to the 
triple-gauge operators $X_{\mu\nu}X^{\nu\lambda}X_{\lambda}^{\mu}$.
They can be found in \cite{Grzadkowski:2010es}. 
All coefficients $c_i$ and $\beta_1$ scale as ${\cal O}(\xi/ 16 \pi^{2})$. 
We used the shorthand notation
\begin{align}
\mathcal{O}_{q}&= 2(\bar{q}\tau_{L}\gamma^{\mu}q) \langle L_{\mu} \tau_{L}\rangle +
(\bar{q}UP_{12}U^{\dagger}\gamma^{\mu}q) \langle P_{21}U^{\dagger}L_{\mu}U\rangle +  
(\bar{q}UP_{21}U^{\dagger}\gamma^{\mu}q) \langle P_{12}U^{\dagger}L_{\mu}U\rangle
\nonumber\\
\mathcal{O}_{l} &= 2(\bar{l}\tau_{L}\gamma^{\mu}l) \langle L_{\mu} \tau_{L}\rangle 
+ (\bar{l}UP_{12}U^{\dagger}\gamma^{\mu}l) \langle P_{21}U^{\dagger}L_{\mu}U\rangle +  
(\bar{l}UP_{21}U^{\dagger}\gamma^{\mu}l) \langle P_{12}U^{\dagger}L_{\mu}U\rangle.
\end{align} 
where $\tau_L=U T_3 U^\dagger$, $P_{12}=T_1 + i T_2$, $P_{21}=P^\dagger_{12}$.

We note that the result of Eq.~(\ref{eq:7}) relies on the assumption that
custodial symmetry and CP are only broken by weak perturbations. Their 
breaking is thus generated by the gauge and Yukawa couplings.
Spurions must then come with an associated weak coupling, which carries
chiral dimension.
It then follows that the $T$ parameter is loop-suppressed and CP violating 
operators can only show up at NNLO. 

In this sense, Eq.~(\ref{eq:7}) is by construction the most general 
next-to-leading-order correction to the chiral electroweak Lagrangian 
close to the decoupling limit, to first order in $\xi$. 
It is therefore a well-defined approach to a systematic derivation 
of the SILH Lagrangian for generic light Higgs scenarios. Notice that picking 
the leading dependence in $\xi$ from the chiral electroweak operators brings 
in a series of correlations between the different coefficients, all of them 
arising from the doublet structure of the Higgs field that emerges in the 
decoupling limit. As already noted, Eq.~(\ref{eq:7}) is written in the broken 
phase and therefore the impact of NLO effects in the LO parameters has been 
taken care of, which results in some operators not being proportional to 
$(v+h)^2$. Apart from this notational aspect, a comparison with the original 
SILH Lagrangian~\cite{Giudice:2007fh} and its recent 
extension~\cite{Contino:2013kra} shows that: (i) since the construction of 
the operators is a purely infrared issue, all model-dependence of the original 
formulation is inessential and can be removed. As a result, the structure of 
the Lagrangian gets simplified and the role of the relevant scales in the 
problem becomes more transparent; (ii) custodial symmetry breaking through
the $T$-parameter comes with an overall coefficient $1/\Lambda^2$, in agreement 
with the discussion in~\cite{Giudice:2007fh};
(iii) in general there is no extra suppression 
of the $B_{\mu\nu}B^{\mu\nu}H^{\dagger}H$ and $G_{\mu\nu}G^{\mu\nu}H^{\dagger}H$ operators
(see the further discussion in Sec. \ref{sec:so5nlo}).

We emphasize that the effective Lagrangian contains also terms of
higher order in $\xi$. By definition, those go beyond the SILH
approximation. However, some of them, related to the Higgs sector, 
come without loop suppression and would typically be more important than
$\xi/16\pi^2$ terms, as long as $\xi\gg 1/16\pi^2$. In practice, to work them
out explicitly, dimension-8 operators would have to be considered.
We note that working with the full electroweak chiral Lagrangian
automatically includes all orders in $\xi$.

In the following section we will give a more detailed account of how 
custodial symmetry breaking is implemented in the EFT. This can be 
done without relying on the UV dynamics and will therefore lead to a number of 
model-independent conclusions.

%%%%%%%%%%%%%%%%%%%%%%%%%%%%%%%%%%%%%%%%%%%%
%   Custodial Symmetry
%%%%%%%%%%%%%%%%%%%%%%%%%%%%%%%%%%%%%%%%%%%%%%%%%%%%%%%%%%%%%%%%%
\section{Custodial symmetry and its breaking}
\label{sec:custodial}

In this section we consider general properties of custodial symmetry
and its violation in the electroweak effective Lagrangian.
The concept of custodial symmetry is well known. We review it here
to provide the proper context for our subsequent general discussion of its
violation by spurions in effective field theory.
 
We assume that the electroweak sector exhibits the spontaneous breaking
of a global symmetry according to the pattern
\begin{equation}\label{su2lrv}
SU(2)_L\otimes SU(2)_R \to SU(2)_V
\end{equation}
The associated Goldstone fields $\varphi^a$, $a=1,2,3$, parametrize the 
coset of the symmetry breaking in (\ref{su2lrv}), expressed through the 
$SU(2)$ matrix field $U=\exp(2i\varphi^a T^a/v)$, where $T^a$ are the 
generators of $SU(2)$. Under $SU(2)_L\otimes SU(2)_R$ the field $U$ transforms
as $U\to g_L U g^\dagger_R$, with $g_{L,R} \in SU(2)_{L,R}$. The vacuum 
$U=\mathbf{1}$
breaks this symmetry but remains invariant under $SU(2)_V$, defined by
$SU(2)$ transformations that obey $g_L=g_R\equiv g_V$.
The residual, global invariance under $SU(2)_V$ is commonly referred to
as the {\it custodial symmetry} \cite{Sikivie:1980hm}.
It is useful to distinguish two somewhat different meanings of this term.
In the narrow sense, custodial symmetry refers only to the spontaneously
broken dynamics itself, that is to the scalar sector (Higgs fields) or
the corresponding new strong interactions. In the general sense,
custodial symmetry refers to all interactions, strong (scalar) dynamics
and weak perturbations (e.g. from gauge or Yukawa couplings).

When (part of) the symmetry $SU(2)_L\otimes SU(2)_R$ is gauged, some of the
gauge fields become massive via the Higgs mechanism. It is instructive to
consider the following possibilities of gauging a subgroup of 
$SU(2)_L\otimes SU(2)_R$ and the resulting spectrum of gauge bosons:
a) $SU(2)_L$ (3 massive, degenerate gauge bosons); 
b) $SU(2)_V$ (3 massless gauge bosons); 
c) $SU(2)_L\otimes SU(2)_R$ (3 massive, degenerate and 3 massless gauge bosons);
d) $SU(2)_L\otimes U(1)_Y$ (Standard Model, 1 massless, 3 massive
gauge bosons with $M_W\not= M_Z$).
By the assumption of (\ref{su2lrv}), all cases have a custodial symmetry
in the narrow sense. In the general sense of the term, custodial symmetry
is violated in the Standard Model ($M_W\not= M_Z$), while
cases a), b) and c) remain custodially symmetric, despite the weak gauging.

The distinction between custodial symmetry in the general or the narrow
sense is of course a matter of definition. However, it clarifies
apparently different uses of the term in the existing literature.
For instance, among the electroweak oblique corrections, the $T$ parameter,
but not the $S$ parameter, is referred to as a measure of custodial 
symmetry breaking in \cite{Barbieri:2004qk}. On the other hand, also the $S$ 
parameter is viewed as a violation of custodial symmetry 
in \cite{Grinstein:1991cd}.
The apparent inconsistency is resolved when the former usage of custodial
symmetry is understood in the narrow sense, the latter in the general sense
of this term. 

In the following, unless stated otherwise, we will adopt the meaning of
custodial symmetry in the general sense as defined above.
Hence, both $M_W\not= M_Z$ and the $S$ parameter violate custodial symmetry,
at leading and next-to-leading order, respectively.

In general, the pattern of explicit breaking of custodial symmetry can
be described by spurions $\omega$. We will prove that, in the context of 
(\ref{su2lrv}), the only spurion of custodial
symmetry breaking in the effective Lagrangian is given by $T^3_R$,
the third generator of $SU(2)_R$. As an illustration of this general theorem
we point out how it is realized in the electroweak chiral Lagrangian at 
leading and next-to-leading order, and also in the usual Standard Model 
with a linearly realized Higgs sector through operators of dimension 6. 

For the case at hand, the spurions are {\it a priori}  general 
$2\times 2$ matrices\footnote{This is because all terms in the Lagrangian are
built from fermion bilinears, $U$ and $h$ fields, gauge field strengths and
covariant derivatives, all of which come with an even number of
$SU(2)_{L,R}$ indices. Invariants can thus only be formed by contracting
with matrices rather than with $SU(2)$ doublets as spurions.}  
with formal transformation properties under
$SU(2)_L\otimes SU(2)_R$, such that invariants under this symmetry can be built,
in general involving also $U$ and further fields. 
The transformation properties reflect the physical origin of a given spurion.
Keeping $\omega$ fixed at 
its true (constant) value breaks the global symmetry in the appropriate way. 

Any spurion must, in general, have one of the three possible 
transformation rules under the global group:
\begin{eqnarray}
\omega &\to& g_L\omega g_R^\dagger \label{chitrafolr}\\
\omega &\to& g_L\omega g_L^\dagger \label{chitrafoll}\\
\omega &\to& g_R\omega g_R^\dagger \label{chitraforr}
\end{eqnarray}
An invariant $\omega\to \omega$ is 
trivial and does not lead to custodial symmetry breaking. 
The hermitian conjugate version of (\ref{chitrafolr}) is understood.

A general $2\times 2$ matrix $\omega$ can be written as a linear combination 
of the unit matrix $\mathbf{1}$ and $T^a$, with complex coefficients.
An essential restriction arises when part of the
global symmetry is gauged, since only spurions consistent with
gauge invariance are allowed. In the case of electroweak theory, the 
entire $SU(2)_L$ and the weak hypercharge subgroup $U(1)_Y$ of $SU(2)_R$
are gauged. A spurion transforming as (\ref{chitrafolr}) would break
local $SU(2)_L$ and is therefore forbidden. Scenario (\ref{chitrafoll})
likewise breaks $SU(2)_L$ unless $\omega\sim \mathbf{1}$, 
which is the trivial case. Similarly, (\ref{chitraforr}) breaks $U(1)_Y$
unless $\omega\sim \mathbf{1}$ or $\omega\sim T^3_R$.
This leaves $T^3_R$ as the only nontrivial spurion and proves our
assertion.

The allowed spurions are different when a different part of the global
group is gauged. An example is the chiral perturbation theory of
pions, where the spontaneous breaking of the global symmetry also follows
(\ref{su2lrv}), gauged under the electromagnetic $U(1)$. The allowed 
spurions are then $\omega\sim \mathbf{1}$ and $\omega\sim T^3$,
each transforming formally under (\ref{chitrafolr}), (\ref{chitrafoll})
or (\ref{chitraforr}). This amounts to the quark mass term transforming
as (\ref{chitrafolr}), and the electric charge operator transforming
as (\ref{chitrafoll}) or (\ref{chitraforr}).

The fact that $T^3_R$ is the only spurion of custodial breaking
under the electroweak gauging of (\ref{su2lrv}), can be illustrated
with concrete examples. Consider first the usual (minimal) Standard Model. 
The SM Lagrangian can be viewed as the low-energy effective theory of any 
general UV completion that might exist. There are two sources of custodial 
symmetry breaking: weak hypercharge gauge interactions, and the difference 
in up- and down-fermion Yukawa couplings. Both are indeed governed by $T^3_R$. 
In order to see that this is not just an accidental feature of the 
lowest-order Lagrangian, one may inspect the full set of dimension-6
operators as classified in \cite{Grzadkowski:2010es}. These can be written 
in terms of the Goldstone matrix $U$ and the Higgs singlet $h$, rather than
in terms of the Higgs doublet $\phi$. This representation has been discussed 
e.g. in \cite{Buchalla:2012qq,Buchalla:2013rka}. In this way it can be
demonstrated explicitly that, again, the only spurion of custodial breaking 
that appears is $T^3_R$. The same observation holds for the electroweak chiral
Lagrangian at leading and next-to-leading order described
in \cite{Buchalla:2013rka}.
Some of the operators in \cite{Buchalla:2012qq,Buchalla:2013rka}
are written in terms of the matrices $P_{12}=T_1 + i T_2$ and
$P_{21}=T_1 - i T_2$. To make the reduction to the spurion $T_3$
manifest one may use the identity
\begin{equation}\label{p12p21}
(P_{12})_{ij} (P_{21})_{kl} =-\frac{1}{4}\delta_{ij}\delta_{kl} +\frac{1}{2}
\delta_{il}\delta_{kj} - (T_3)_{ij} (T_3)_{kl}
+\frac{1}{2} (T_3)_{il} \delta_{kj} - \frac{1}{2} \delta_{il} (T_3)_{kj}
\end{equation} 

The discussion of this section demonstrates in particular
that the presence of $T^3_R$ as the only spurion
of custodial-symmetry breaking in the general electroweak chiral 
Lagrangian is a fully general,  model-independent property
of the effective field theory formulation, in contrast
to the claims in \cite{Contino:2013kra}.

%%%%%%%%%%%%%%%%%%%%%%%%%%%%%%%%%%%%%%%%%%%%%%%%%%%%%%%%%%%%%%%%%
%   SO(5) at NLO
%%%%%%%%%%%%%%%%%%%%%%%%%%%%%%%%%%%%%%%%%%%%%%%%%%%%%%%%%%%%%%%%%
\section{\boldmath $SO(5)/SO(4)$ model at NLO in the 
chiral \\ expansion}
\label{sec:so5nlo}

We now would like to show how some of the features that we discussed
arise in the context of a specific model, the minimal composite Higgs
model of \cite{Agashe:2004rs,Contino:2006qr}.
This model assumes spontaneous symmetry breaking of $SO(5)$ down to $SO(4)$ 
at a scale $f$, which generates four Goldstone bosons. They span the
coset space and can be parametrized as \cite{Contino:2010rs} 
(see Appendix for details)
\begin{align}\label{sigmahu}
\Sigma(h_{\hat a})=\left(\begin{array}{c}
\displaystyle\frac{s}{2}\langle U\lambda_{\hat a}^{\dagger}\rangle 
\\ c
\end{array}\right),\qquad \lambda_{\hat a}=(i{\vec{\sigma}},1_2),\qquad
s=\sin\frac{\ch}{f},\qquad c=\cos\frac{\ch}{f}
\end{align}
Above we used the fact that $SO(4)$ is isomorphic to $SU(2)_L\otimes SU(2)_R$ 
to express the $SO(4)$ vector $h_{\hat a}$ in terms of the 
$SU(2)_L\otimes SU(2)_R$ bifundamental field $U$ and the $h_{\hat a}$ 
modulus $\ch$. The custodial-preserving 
$SU(2)_L\otimes SU(2)_R$ is further broken (explicitly) by the couplings to 
gauge bosons and fermions. The spurion for this breaking is $t_3^R$ and is 
accompanied by powers of $g^{\prime}$ and/or Yukawa couplings $y_f$. 
For simplicity, in the following we will set fermions aside and focus on 
the CP-even bosonic sector. 

The leading-order Lagrangian (chiral dimension $\chi=2$) takes the form
\begin{align}\label{LO}
{\cal{L}}&=\frac{f^2}{2}\Sigma_{\mu}^T\Sigma^{\mu} - V=
\frac{1}{2}\partial_{\mu}\ch\partial^{\mu}\ch+
\frac{f^2}{4}\langle L_{\mu}L^{\mu}\rangle s^2 - V
\end{align}
where $\Sigma_\mu\equiv D_\mu\Sigma$.
The gauge kinetic terms are understood.
The leading-order potential is
\begin{equation}\label{vab}
V=\alpha\, \Sigma^T n -4\beta\,\Sigma^T t_3^Rt_3^R \Sigma
=\alpha\, c -\beta\, s^2
\end{equation}
where $n=(0,0,0,0,1)^T$ and $t^R_3$ are the $SO(5)$-breaking spurions
that are consistent with SM gauge invariance. The vector $n$ conserves
custodial symmetry, the matrix $t^R_3$ violates it. Both are
related through $n n^T=1-4 t^R_3 t^R_3$.

The coefficients have $\chi=2$ since they are loop-suppressed and
they scale as $\alpha$, $\beta\sim f^4$. A realization of such a potential
in a specific model has been discussed e.g. in \cite{Contino:2010rs}.

We note that the two terms in (\ref{vab}) are given by the two independent
expressions that can be built at leading order from the spurions of
$SO(5)$ breaking, $n$ and $t^R_3$.
For $\beta >0$ and $|\alpha|\leq 2\beta$, the potential in (\ref{vab})
exhibits spontaneous symmetry breaking, generating the vacuum expectation
value $\langle \ch \rangle$ via
\begin{align}
\xi=\frac{v^2}{f^2}=\sin^2\frac{\langle \ch \rangle}{f} = 
1-\left(\frac{\alpha}{2\beta}\right)^2,
\qquad  v<\langle \ch\rangle<\frac{\pi}{2}v
\end{align}
where $\langle \ch \rangle$ ranges from the decoupling to the nondecoupling 
limit. The resulting mass of the physical scalar boson 
$h\equiv\ch - \langle\ch\rangle$ is
\begin{equation}\label{mh2}
m^2_h=\frac{2\beta\xi}{f^2}={\cal O}(v^2)
\end{equation}

To construct the operators at NLO ($\chi=4$), it is necessary to 
employ the general method of Callan, Coleman, Wess and 
Zumino~\cite{Coleman:1969sm,Callan:1969sn}. For the case of the $SO(5)/SO(4)$ 
coset this has been performed in great detail in~\cite{Contino:2011np}
(see \cite{Alonso:2014wta} for a recent discussion of this and other cosets). 
Here we restrict ourselves to quoting the main results, adding some comments
and discussing the matching to the electroweak chiral Lagrangian at scale $v$.

One defines $d_\mu$ and $E_\mu$ through~\cite{Contino:2011np} 
\begin{equation}\label{caludefde}
-i{\cal U}^\dagger D_\mu {\cal U} = d^{\hat a}_\mu t^{\hat a} + E^a_\mu t^a
\equiv d_\mu + E_\mu
\end{equation}
Here ${\cal U}=\exp(\sqrt{2}it^{\hat a} h_{\hat a}/f)$ and $t^{\hat a}$ ($t^a$)
are the broken (unbroken) generators of $SO(5)\to SO(4)$.
$D_\mu=\partial_\mu + i A_\mu$ is the covariant derivative with
$A_\mu=A^{\hat a}_\mu t^{\hat a} + A^a_\mu t^a$ in the most general case.
(Here the coupling has been absorbed in $A_\mu$).
In practice, we will be mostly interested in gauging the standard-model
group, in which case $A^{\hat a}_\mu=0$.
The following building blocks are useful~\cite{Contino:2011np}:
\begin{equation}\label{eler}
\partial_\mu E_\nu - \partial_\nu E_\mu +i[E_\mu,E_\nu]\equiv
E_{\mu\nu}\equiv E^L_{\mu\nu} + E^R_{\mu\nu}
\end{equation}  
\begin{equation}\label{fmflfr}
f_{\mu\nu}={\cal U}^\dagger F_{\mu\nu} {\cal U}\equiv 
f^-_{\mu\nu} + f^L_{\mu\nu} + f^R_{\mu\nu}
\end{equation}
Here $f^-_{\mu\nu}\equiv f^{-,\hat a}_{\mu\nu} t_{\hat a}$,
$f^L_{\mu\nu}\equiv f^{L,a}_{\mu\nu} t^L_a$, $f^R_{\mu\nu}\equiv f^{R,a}_{\mu\nu} t^R_a$, 
and similarly for $E^{L,R}_{\mu\nu}$, where the six unbroken generators $t^a$ 
are decomposed into the generators $t^{L,R}_a$of $SU(2)_{L,R}$. 
$F_{\mu\nu}$ is the field strength of $A_\mu$.  

The kinetic term in (\ref{LO}) can then be written as
\begin{equation}\label{loccwz}
{\cal L}=\frac{f^2}{2}\Sigma_{\mu}^T\Sigma^{\mu}
\equiv \frac{f^2}{4} \langle d_\mu d^\mu \rangle
\end{equation}

The NLO operators can be constructed from the
building blocks above. The CP even operators read  \cite{Contino:2011np} 
\begin{align}\label{o15m}
O_1 &= \langle d_\mu d^\mu\rangle^2 \nonumber\\
O_2 &= \langle d_\mu d_\nu\rangle \langle d^\mu d^\nu\rangle\nonumber\\
O_3 &= \langle E^L_{\mu\nu} E^{L,\mu\nu}\rangle -
       \langle E^R_{\mu\nu} E^{R,\mu\nu}\rangle\nonumber\\
O^+_4 &= \langle (f^L_{\mu\nu} + f^R_{\mu\nu}) i[d^\mu, d^\nu]\rangle\nonumber\\
O^+_5 &= \langle (f^-_{\mu\nu})^2\rangle\nonumber\\
O^-_4 &= \langle (f^L_{\mu\nu} - f^R_{\mu\nu}) i[d^\mu, d^\nu]\rangle\nonumber\\
O^-_5 &= \langle (f^L_{\mu\nu})^2 - (f^R_{\mu\nu})^2\rangle
\end{align}

We remark that the operator $O_3$ in this list is redundant:
\begin{equation}\label{o345}
O_3 = O^-_5 - 2 O^-_4
\end{equation}
This was also noted recently in \cite{Alonso:2014wta}.
The remaining operators can be expressed in terms of the $2\times 2$
Goldstone field $U$ and the Higgs singlet $|h|$ of the chiral Lagrangian
based on the coset $SU(2)_L \otimes SU(2)_R/SU(2)_V$.
We find
\begin{align}\label{olchi}
O_1 &= \left(\frac{2}{f^2} \partial_\mu |h| \partial^\mu |h|
 + s^2 \langle L_\mu L^\mu\rangle\right)^2 \nonumber\\
O_2 &= \left(\frac{2}{f^2} \partial_\mu |h| \partial_\nu |h|
 + s^2 \langle L_\mu L_\nu\rangle\right)^2 \nonumber\\
O^+_4 &= -s^2 \langle g D_\mu W^{\mu\nu} L_\nu -
        g^\prime\partial_\mu B^{\mu\nu} \tau_L L_\nu 
  +\frac{g^2}{2}(W_{\mu\nu})^2 +\frac{g^{\prime 2}}{2}(B_{\mu\nu} T_3)^2 
  - g^\prime g B_{\mu\nu} W^{\mu\nu}\tau_L\rangle\nonumber\\ 
O^+_5 &= s^2\langle g^2 (W_{\mu\nu})^2 + g^{\prime 2}(B_{\mu\nu} T_3)^2 
         - 2  g^\prime g B_{\mu\nu} W^{\mu\nu}\tau_L\rangle \nonumber\\
O^-_4 &=i\frac{c}{2}(s^2+2) \langle g W_{\mu\nu} [L^\mu,L^\nu] -
       g^\prime B_{\mu\nu} \tau_L [L^\mu,L^\nu]\rangle \nonumber\\  
  &+ 2c \langle g D_\mu W^{\mu\nu} L_\nu +g^\prime\partial_\mu B^{\mu\nu} \tau_L L_\nu 
 +\frac{g^2}{2}(W_{\mu\nu})^2 -\frac{g^{\prime 2}}{2}(B_{\mu\nu} T_3)^2 \rangle 
\nonumber\\
O^-_5 &= 2 c\langle g^2 (W_{\mu\nu})^2 - g^{\prime 2}(B_{\mu\nu} T_3)^2 \rangle
\end{align}
Here the gauging has been restricted to the standard-model group
and the couplings have been factored out of the gauge fields.
The terms with $D_\mu W^{\mu\nu}$ and $\partial_\mu B^{\mu\nu}$ in 
$O^\pm_4$ are reducible upon using the equations of motion.   
Obviously, $O^+_{4,5}\to 0$ in the limit $|h|\to 0$. Note that in the same 
limit $O^-_4$ also vanishes upon integrating by parts, while $O^-_5$ just 
renormalizes the gauge kinetic terms.

The operators on the r.h.s. of (\ref{olchi}) match
the electroweak chiral Lagrangian in the basis 
of~\cite{Buchalla:2013rka} after eliminating redundant terms and expanding
around the Higgs vacuum expectation value. Indeed, $O_{1,2}$ correspond to 
${\cal O}_{Di}$ in~\cite{Buchalla:2013rka} with $i=1,2,7,8,11$, whereas 
$O^\pm_{4,5}$ contain ${\cal O}_{Xh1,2}$ and ${\cal O}_{XU1,7,8}$. 

The $SO(5)/SO(4)$ example illustrates how its effective-theory formulation
can be expressed in terms of the general chiral Lagrangian   
of~\cite{Buchalla:2013rka}. 
Expanding the former to first order in $\xi$ provides an explicit
realization of the SILH Lagrangian derived in Section \ref{sec:silh}. 
It also exhibits the presence of $T^3_R$
as the only spurion of custodial symmetry breaking, in agreement
with the theorem of Section \ref{sec:custodial}. 

The representation of the operators in (\ref{olchi}) makes it
explicit that the Higgs couples to a pair of field-strength factors
only in the combinations 
\begin{align}\label{hxx}
& \langle g^2 (W_{\mu\nu})^2 + g^{\prime 2}(B_{\mu\nu} T_3)^2 
         - 2 g^\prime g B_{\mu\nu} W^{\mu\nu}\tau_L\rangle \nonumber\\
& \langle g^2 (W_{\mu\nu})^2 - g^{\prime 2}(B_{\mu\nu} T_3)^2 \rangle
\end{align}  
These do not contain the photon-photon component $F_{\mu\nu}F^{\mu\nu}$
and hence there is no $h\to\gamma\gamma$ operator at this order.
This has been emphasized in \cite{Giudice:2007fh} and explained
as the consequence of a residual shift symmetry that commutes with
the electric charge $Q$, similar to the absence of 
$(\pi^0)^2 F_{\mu\nu}F^{\mu\nu}$ at NLO in chiral perturbation 
theory \cite{Terentev:1972ix,Donoghue:1988eea}.
This feature is valid for the nonlinear (bosonic) Lagrangian
defined at scale $f$ and represented at NLO through the terms
in (\ref{olchi}). However, the electroweak effective Lagrangian
is defined at the scale $v$. In the limit of small $\xi=v^2/f^2$
the physics at scale $f$ is then integrated out, which may induce
the local operator $h F_{\mu\nu}F^{\mu\nu}$ with a coefficient
of order $\xi/16\pi^2$ as in~(\ref{eq:7}), that is in general without
extra suppression. The same holds for the coupling to gluons,
$h G_{\mu\nu}G^{\mu\nu}$.
An example is provided by fermion representations
in minimal composite Higgs models, which induce local $h\to\gamma\gamma$
and $h\to gg$ operators with coefficients of size $\xi/16\pi^2$
\cite{Falkowski:2007hz,Carena:2014ria}.
This is due to an explicit soft breaking of $SO(5)$ in the fermionic
sector at scale $f$.

%%%%%%%%%%%%%%%%%%%%%%%%%%%%%%%%%%%%%%%%%%%%%%%%%%%%%%%%%%%%%%%%%          
%   Conclusions                                                           
%%%%%%%%%%%%%%%%%%%%%%%%%%%%%%%%%%%%%%%%%%%%%%%%%%%%%%%%%%%%%%%%%          
\section{Conclusions}
\label{sec:concl}

Strongly-coupled scenarios are viable candidates to explain the mechanism of 
electroweak symmetry breaking. In their minimal version, one assumes a 
$SU(2)_L\otimes SU(2)_R\to SU(2)_V$ breaking pattern at a scale $v$, 
with a light Higgs (presumably, but not necessarily, a pseudo-Goldstone boson) 
and new physics starting around 
the TeV scale. Under these assumptions, the most efficient description of 
the physics at present-day colliders is provided by the electroweak chiral 
Lagrangian. In this paper we have explicitly shown that the so-called 
SILH Lagrangian can be recovered as a special limit of the latter. 
To the best of 
our knowledge, this viewpoint offers the first rigorous derivation of SILH 
and helps to clarify some of its aspects, especially those related with power 
counting and the breaking of custodial and shift symmetries. 

Our main conclusions can be summarized in the following points:
\begin{itemize} 
\item We emphasize that the small-$\xi$ limit of the electroweak chiral
Lagrangian relies on a double expansion in both, powers of $\xi$ and the 
number of loops. Phenomenologically, terms of order $\xi^2$ might be larger
than the $\xi/16\pi^2$ terms included in the conventional SILH Lagrangian.
The electroweak chiral Lagrangian represents the resummation to all
orders in $\xi$. 
\item The SILH Lagrangian can be understood as the electroweak chiral 
Lagrangian to first order in $\xi=v^2/f^2$. This allows a systematic 
construction of the effective-theory operators with a well-defined 
power counting 
and without relying on particular UV completions. The resulting set of 
dimension-6 operators comes from both the LO and NLO chiral Lagrangian 
and they are suppressed, respectively, by $1/f^2$ and $1/\Lambda^2$. 
\item In scenarios where the Higgs is a pseudo-Goldstone boson, the 
$h\to \gamma\gamma$ and $h\to gg$ amplitudes at scale $v$ receive local 
contributions of order $\xi/16\pi^2$. These can arise from integrating out
new states at scale $f$, which may exist in realistic models.
\item We prove that, given the $SU(2)_L\otimes SU(2)_R\to SU(2)_V$ breaking 
pattern, custodial symmetry breaking is described by a single spurion, 
namely $T^3_R$. If custodial symmetry is assumed to be preserved by the 
strong sector, and only broken explicitly by the weak sector (gauge and 
Yukawa couplings), the $T$-parameter appears as a NLO effect 
(of chiral dimension 4) and comes with a suppression of $1/\Lambda^2\ll 1/f^2$.
\item As a concrete illustration of the previous points, we have considered 
the NLO operators of the CP-even bosonic sector of the $SO(5)/SO(4)$ model and 
matched them to the electroweak chiral Lagrangian. 
\end{itemize}
To summarize, the electroweak chiral Lagrangian, formulated 
with the vacuum misalignment parameter $\xi$, gives not only a 
description of strict nondecoupling scenarios ($\xi\sim 1$), but it is also 
valid for softly nondecoupling constructions ($\xi\ll 1$), like the 
SILH Lagrangian. The electroweak chiral Lagrangian thus provides the 
well-defined starting point for the construction of generic EFT descriptions 
of electroweak physics with a strong sector. 
Importantly, in the small-$\xi$ limit, the electroweak chiral Lagrangian
implies a pattern for the coefficients of dimension-6 operators
characteristic of a strongly-interacting Higgs sector, which can be tested
against experiment.

%%%%%%%%%%%%%%%%%%%%%%%%%%%%%%%%%%%%%%%%%%%%%%%%%%%%%%%%%%%%%%%%%
%     Acknowledgements
%%%%%%%%%%%%%%%%%%%%%%%%%%%%%%%%%%%%%%%%%%%%%%%%%%%%%%%%%%%%%%%%%
\section*{Acknowledgements}

We thank Michael Trott for interesting discussions on custodial symmetry and
Gian Giudice, Christophe Grojean, Alex Pomarol and Riccardo Rattazzi for 
comments on a preliminary version of the manuscript. This work was performed 
in the context of the ERC Advanced Grant project `FLAVOUR' (267104) and was 
supported in part by the DFG cluster of excellence `Origin and Structure 
of the Universe' and DFG grant BU 1391/2-1.

\appendix
\numberwithin{equation}{section} 
%%%%%%%%%%%%%%%%%%%%%%%%%%%%%%%%%%%%%%%%%%%%%%%%%%%%%%%%%%%%%%%%%
%   Appendix
%%%%%%%%%%%%%%%%%%%%%%%%%%%%%%%%%%%%%%%%%%%%%%%%%%%%%%%%%%%%%%%%%
\section{\boldmath $SO(5)/SO(4)$ Goldstone field}
\label{sec:appso5}

In this Appendix we collect some technical details used in
the discussion of Section~\ref{sec:so5nlo}.

When $SO(5)$ is spontaneously broken to $SO(4)$, the Goldstone 
multiplet can be pa\-ra\-me\-trized by (see e.g. \cite{Contino:2010rs})
\begin{align}\label{sigmaha}
\Sigma(h_{\hat a})={\cal U}\, \Sigma_0,\qquad 
\Sigma_0=\left(\begin{array}{c}
0_4\\1
\end{array}\right)
\end{align}
where 
\begin{equation}\label{caludef}
{\cal U}=\exp(\sqrt{2}it^{\hat a} h_{\hat a}/f)
\end{equation}
with $h_{\hat a}$ ($\hat a =1,\ldots, 4$) a $SO(4)$ vector and $t^{\hat a}$ 
the broken generators that span the 4-parameter coset. 
Using the realization ($i,j=1,\ldots, 5$)
\begin{align}\label{taij}
t^{\hat a}_{ij}=-\frac{i}{\sqrt{2}}
\left(\delta^{\hat a}_i \delta^5_j - \delta^{\hat a}_j \delta^5_i\right)
\end{align}
direct substitution of the generators above yields
($s=\sin{\ch/f}$, $c=\cos{\ch/f}$)
\begin{align}
\Sigma(h_{\hat a})=\left(\begin{array}{c}
\hat{h}_{\hat a} s \\ c
\end{array}\right),\qquad \hat{h}_{\hat a}=\frac{h_{\hat a}}{\ch},
\qquad \ch=\sqrt{h_{\hat a} h_{\hat a}}
\end{align}
Since $SO(4)$ is isomorphic to $SU(2)_L\otimes SU(2)_R$, one can relate the 
$SO(4)$ vector to a complex $SU(2)_L\otimes SU(2)_R$ bidoublet $H$ and its 
polar decomposition into $\ch$ and the $SU(2)$ matrix $U$,
\begin{align}
H=(\tilde\phi, \phi)=\left(\begin{array}{cc}
h_4+ih_3&h_2+ih_1\\
-(h_2-ih_1)&h_4-ih_3
\end{array}\right)
=h_{\hat a}\lambda_{\hat a}\equiv \ch U,\qquad \lambda_{\hat a}=(i{\vec{\sigma}},1_2)
\end{align}
which implies
\begin{align}\label{hhatu}
\hat{h}_{\hat a}=\frac{1}{2}\langle U\lambda_{\hat a}^{\dagger}\rangle
\end{align}
The doublet $\phi$ corresponds to the SM Higgs. 
The present definitions ensure that $\phi$ transforms as a $SU(2)_L$ doublet 
with weak hypercharge $Y=\frac{1}{2}$, if the $SO(4)$ generators 
are realized as  
\begin{align}
t_1^L&=\frac{i}{2}\left(\begin{array}{cc}
0 &-\sigma_1\\
\sigma_1 & 0
\end{array}\right);\qquad 
t_2^L=\frac{i}{2}\left(\begin{array}{cc}
0 &\sigma_3\\
-\sigma_3 & 0
\end{array}\right);\qquad 
t_3^L=\frac{i}{2}\left(\begin{array}{cc}
-i\sigma_2 &0\\
0 & -i\sigma_2
\end{array}\right)\nonumber\\
t_1^R&=\frac{i}{2}\left(\begin{array}{cc}
0 &i\sigma_2\\
i\sigma_2 &0
\end{array}\right);\qquad 
t_2^R=\frac{i}{2}\left(\begin{array}{cc}
0 &1\\
-1 &0
\end{array}\right);\qquad
t_3^R=\frac{i}{2}\left(\begin{array}{cc}
-i\sigma_2 & 0\\
0 & i\sigma_2
\end{array}\right)\label{tlri} 
\end{align}
which satisfy
\begin{align}
t_a^{L,R}t_b^{L,R}&=\frac{1}{4}\delta_{ab}+\frac{i}{2}\varepsilon_{abc}t_c^{L,R};
\qquad [t_a^L,t_b^R]=0
\end{align}
The full set of $SO(5)$ generators is then given by (\ref{taij}) 
and the obvious extension of (\ref{tlri}) 
to $5\times 5$ matrices \cite{Contino:2010rs}.

The operators of the $SO(5)/SO(4)$ chiral Lagrangian can be
constructed from the building blocks quoted in Sec. \ref{sec:so5nlo},
taking into account their chiral dimension:
\begin{equation}\label{chidimcalu}
[d_\mu]_c = 1,\qquad [E^{L,R}_{\mu\nu}]_c=[f^{-,L,R}_{\mu\nu}]_c=2
\end{equation}

The operators in terms of ${\cal U}$ can be expressed through
$\ch$ and $U$ using (\ref{hhatu}),
\begin{equation}\label{calubloc}
{\cal U} = \left(\begin{array}{c|c}
1-(1-c)\hat h \hat h^T & s \hat h \\
\hline
-s \hat h^T & c
\end{array}\right)
\end{equation}
and the relations
\begin{equation}\label{lldelta}
\lambda_{ij}^{\hat a}\lambda_{kl}^{\hat a\dagger}=2\delta_{il}\delta_{kj}
\end{equation}
\begin{equation}\label{tarep}
t^L_{a,\hat a \hat b} \langle U\lambda^\dagger_{\hat b}\rangle =
\langle T^a U\lambda^\dagger_{\hat a}\rangle,
\qquad
t^R_{3,\hat a\hat b} \langle U\lambda^\dagger_{\hat b}\rangle=
-\langle U T_3\lambda^\dagger_{\hat a}\rangle
\end{equation}
The resulting operators with chiral dimension 4 are collected in 
Section \ref{sec:so5nlo}.

%%%%%%%%%%%%%%%%%%%%%%%%%%%%%%%%%%%%%%%%%%%%%%%%%%%%%%%%%%%%%%%%%
%     References
%%%%%%%%%%%%%%%%%%%%%%%%%%%%%%%%%%%%%%%%%%%%%%%%%%%%%%%%%%%%%%%%%

\end{document}